\documentclass[12pt]{iopart}
\usepackage{iopams}  
\usepackage{graphicx}
\usepackage{dcolumn}
\usepackage{bm}
\begin{document}
\title[Elastic-like Collision of Gap Solitons in Bragg Gap Regions]{Elastic-like Collision of Gap Solitons in Bragg Gap Regions within Nonlocal Nonlinear Photonic Crystals}

\author{YuanYao Lin, I-Hong Chen, and Ray-Kuang Lee}
\address{Institute of Photonics Technologies, National Tsing-Hua University, Hsinchu 300, Taiwan}
\ead{rklee@ee.nthu.edu.tw}

%\author{YuanYao Lin}
%\affiliation{Institute of Photonics Technologies, National Tsing-Hua University, Hsinchu 300, Taiwan}
%\author{I-Hong Chen}
%\affiliation{Institute of Photonics Technologies, National Tsing-Hua University, Hsinchu 300, Taiwan}
%\author{Ray-Kuang Lee}
%\affiliation{Institute of Photonics Technologies, National Tsing-Hua University, Hsinchu 300, Taiwan}
%\email{rklee@ee.nthu.edu.tw}

\begin{abstract}
We analyze the existence, stability, and mobility of gap solitons in a periodic photonic structure with nonlocal nonlinearity. 
Within the Bragg region of band gaps, gap solitons exhibit better stability and higher mobility due to the combinations of non-locality effect and the oscillation nature of Bloch waves.
Using linear stability analysis and calculating the Peierls-Nabarro potentials, we demonstrate that gap solitons can revive a non-trivial elastic-like collision even in the periodic systems with the help of nonlocal nonlinearity.
Such interesting behaviors of gap solitons in nonlocal nonlinear photonic crystals are believed to be useful in optical switching devices.
\end{abstract} 

\pacs{42.65.Tg, 42.65.Jx, 42.65.Wi}
\vspace{2pc}
\noindent{\it Keywords}: solitons, photonic crystals, nonlocal effect\\
\submitto{\JOA\\ Special Issue featuring selected papers from Optical MEMS and Nanophotonics 2007}
\maketitle
\date{\today}

\section{Introduction}
Solitary waves are self-guided wave packets as they propagate in nonlinear media, remaining localized and preserving their own shape.
However, they are dramatically altered when collide with one another. 
Strictly speaking, only the special case of Kerr nonlinearity is integrable by the inverse scattering transform method.
{\it Soliton} belongs to a special family of solitary waves that are unaffected by collisions.
They are particle-like wave packets supported by the nonlinear action and their collision behavior are strongly depends upon the relative phase between the interacting solitons \cite{book}.
With such asymptotically linear superposition of nonlinear interaction, a new window of soliton-based photonics can be employed in different optical switching devices and communication systems \cite{soliton-ph}.

During last decade, photonic crystals, artificial periodic structures with the modulation in the refractive index, provide an efficient control of wave transmission and localization, making it possible to tailor dispersion, diffraction, and emission of electromagnetic waves \cite{Joannopoulos}.
Combination of Kerr nonlinear material and photonic crystals, nonlinear photonic crystals have revealed a wealth of nonlinear optical phenomena and, in particular, self-trapped nonlinear localized modes in the form of so called gap solitons \cite{Sterke94, Mingaleev, npc-book}.
Gap solitons are unique solutions which can be formed no matter in focusing or defocusing media due to the change of the dispersion/diffraction effect caused by the photonic lattices \cite{Ostrovskaya03}.
Current technology in reconfigurable optical lattices, such as photorefractive crystals \cite{Efremidis2002} and nematic liquid crystals \cite{Peccianti2002}, also pave a new way to control solitary waves by varying the lattice depth and period. 

With both benefits from photonic crystals and solitons, gap soliton is believed to be an important key footstone in soliton-driven photonics.
For Kerr-type nonlinear photonic crystals, the bifurcation and stability of gap solitons in internal reflection and Bragg gap regions are studied with local nonlinear response \cite{Pelinovsky2004}.
However, the limited mobility of gap soliton in the transverse directions due to corresponding lattice potentials draws great concerns for various switching and routing operations \cite{Crist2003, kartashov2004}.
Recently it has been predicted that with {\it non-locality} solitons can move across the lattice drastically easily in the internal reflection region \cite{Torner-gap}.
Nonlocal effect comes to play an important role as the characteristic response function of the medium is comparable to the transverse content of the wave packet \cite{Wieslaw2000}. 
Experimental observations of nonlocal response also have been demonstrated in various systems, such as photorefractive crystals \cite{Duree93}, nematic liquid crystals \cite{Conti03}, and thermo-optical materials \cite{Rotschild05}.
The study of nonlocal nonlinearity brings new features in solitons \cite{Snyder97}, such as modification of modulation instability \cite{Krolikowski04}, azimuthal instability \cite{Anton06}, and transverse instability \cite{TI-YY}.
Suppression of collapse in multidimensional solitons \cite{Bang02}, change of the soliton interaction \cite{Peccianti02},  formation of soliton bound states \cite{Torner05} and unique families of dark-bright soliton pairs \cite{nlocal-YY} are also predicted recently.

For nonlocal nonlinear medium, the non-locality is known to improve the stabilization of solitons due to the diffusion mechanism of nonlinearity.
Nevertheless, to the best of our knowledge, the existences, stabilities, mobilities, and collisions of gap solitons in Bragg gap regions have not been reported. 
In this work, we extend the concept of gap solitons in nonlocal nonlinear photonic crystals \cite{Torner-gap}, from internal reflection to Bragg gap regions.
The propagation and stability of gap solitons with an imprinted transverse index modulation under the influence of nonlocal effect are studied.
With the oscillation nature of wave packets in the Bragg regions, we show that gap soliton are more stable and movable even with a small degree of non-locality.

This work is organized as follows, first we show that in the band gap region nonlinear Bloch wave can support bright soliton solutions.
Families of even and odd modes of bright gap solitons imprinted onto Bloch wave in local and nonlocal non-linearities are found numerically.
Then the modulation instability of these nonlocal gap soliton families are analyzed by a standard linear stability analysis, and the Peierls-Nabarro (PN) potentials that inhibit the mobility of the gap solitons are also calculated in terms of nonlocality.  
Finally, we address the transverse mobility and soliton interactions under the influence of non-locality with the presence of periodic potentials.
Based on the dramatic reduction of the PN potential barrier for gap solitons in Bragg regions, we demonstrate a non-trivial elastic-like collision between gap solitons, which should be useful for optical switching devices based on soliton collisions.
\section{Nonlocal solitons in Bragg gaps}
We consider a wave packet propagating along the $z$ axis in the nonlocal nonlinear photonic crystals with a Kerr-type nonlinearity and an exponential-type nonlocal response, which can be modeled by the modified nonlinear Schr{\"o}dinger equation,
\begin{eqnarray}
&&i \frac{\partial \Psi}{\partial z}+\frac{1}{2}\frac{\partial^2 }{\partial x^2}\Psi-V(x)\Psi+n(x, z) \Psi=0,
\label{eqGP1}\\
&&n - d \frac{\partial^2 n}{\partial x^2} =|\Psi|^2, 
\label{eqGPn}
\end{eqnarray}
where $\Psi$ is the envelope function of the wave packet, $x$ is the transverse coordinate, and $n(x, z)$ is the refractive index profile induced by the exponential-type kernel function responding to the intensity soliton intensity \cite{Kr2001}.
$V(x)$ is the periodic potential provided externally in the transverse direction.
The coefficient $d$ stands for the degree of nonlocality which governs the diffusion strength of refractive index.

\begin{figure}
\includegraphics[width=8.0cm]{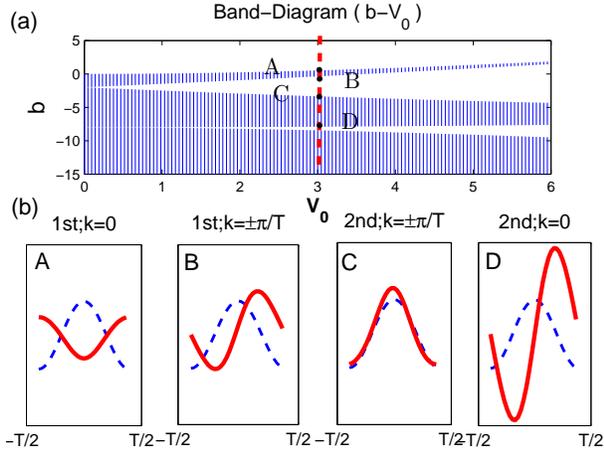}
\caption{(a)The band-gap spectrum of Bloch wave at band edge in linear region for different wave vector $b$ and potential depth $V_0$.
The shaded area shows the allowed band for longitudinal wave vector $b$.
(b)Solid-lines correspond to wave functions of the Bloch state at the band edges, as marked with A, B, C, and D in (a); while dashed-lines indicate periodic potentials, $V(x)=-V_0 cos(2\pi x/T)$.}
\label{Fig:F1}
\end{figure}
\begin{figure}
\includegraphics[width=8.0cm]{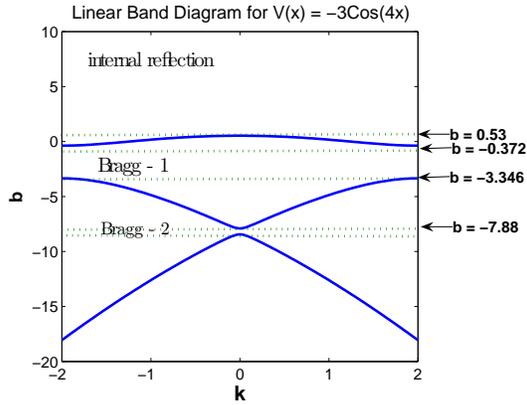}
\caption{The linear band-gap diagram for the longitudinal and transverse wave vectors, $b-k$, with the periodic function $V(x)= - 3cos(4x)$ is used. Different gap regions are marked as internal reflection and Bragg gap regions.}
\label{Fig:F2}
\end{figure}
\subsection{Linear band diagram}
The periodicity of potential $V(x)$ in Eq.(\ref{eqGP1}) suggests that the stationary states can be expanded by Bloch waves $ \Psi(x, z) = f(x) e^{ikx + ibz} $, where $f(x+T)=f(x)$ is a periodic function with period $T$.
If the medium is linear, $n(x,z) = 0$, we can drop the nonlinear index response and rewrite Eq.(\ref{eqGP1}) in terms of $f(x)$,
\begin{eqnarray}
\left(\frac{1}{2}\frac{d^2}{dx^2} - \frac{k^2}{2} + ik\frac{d}{dx} -V(x) \right) f = bf,
\label{EqGPL}
\end{eqnarray}
where $k$ is the transverse wave vector of the wave packet and $b$ is the longitudinal wave vector. 
The linear wave spectrum consists of bands of eigenvalues $b_{n,k}$ in which $k(b)$ is a real wave number of the amplitude-bounded oscillatory Bloch waves. 
The bands are separated by gaps where the wave functions are not stationary with $Im(b)\neq 0$. 
In the absence of nonlinearity, the solution at the band edge is exactly periodic stationary Bloch waves; however, in the presence Kerr nonlinearity, bright gap solitons arise from the forbidden gaps in linear case.
Figure \ref{Fig:F1}(a) shows the linear band-gap diagram on the plane ($b$,$V_0$) that is obtained by solving the linear eigenvalue problem in Eq.(\ref{EqGPL}). 
Different Bloch wave patterns at different band edge are plotted in Fig. \ref{Fig:F1}(b) for a comparison. 
In the numerical calculations, we employ $V(x) = -3cos(4x)$ as an example and the corresponding dispersion relation is depicted in Fig. \ref{Fig:F2} where the value of longitudinal wave vector $b$ for each band edge is addressed.
Based on our definitions, a semi-infinite internal reflection region exists for $b > 0$ while finite Bragg gap regions exist for $b < 0$, as indicated in Fig.\ref{Fig:F2}.

\begin{figure}
\includegraphics[width=3.2in]{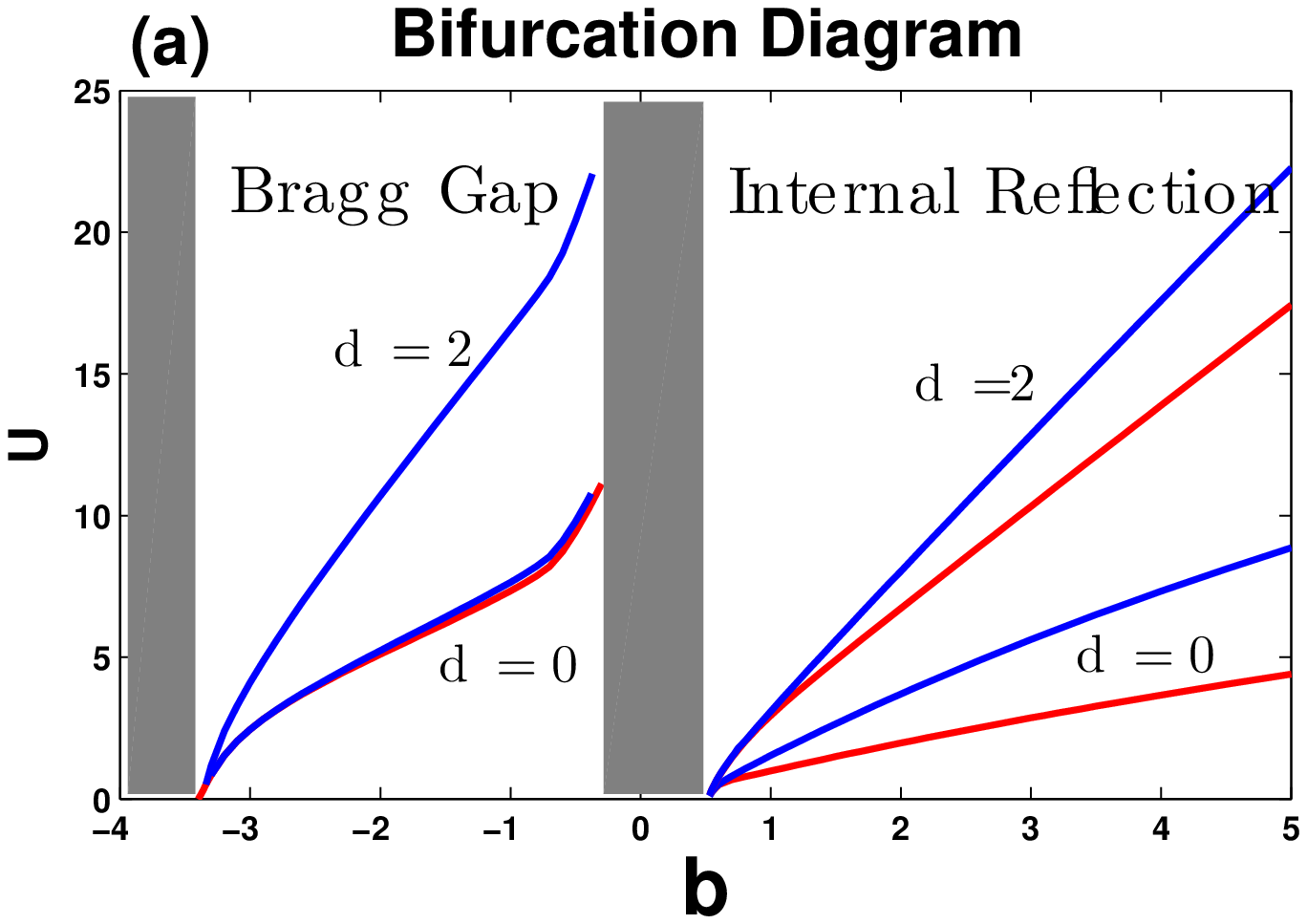}
\includegraphics[width=3.6in]{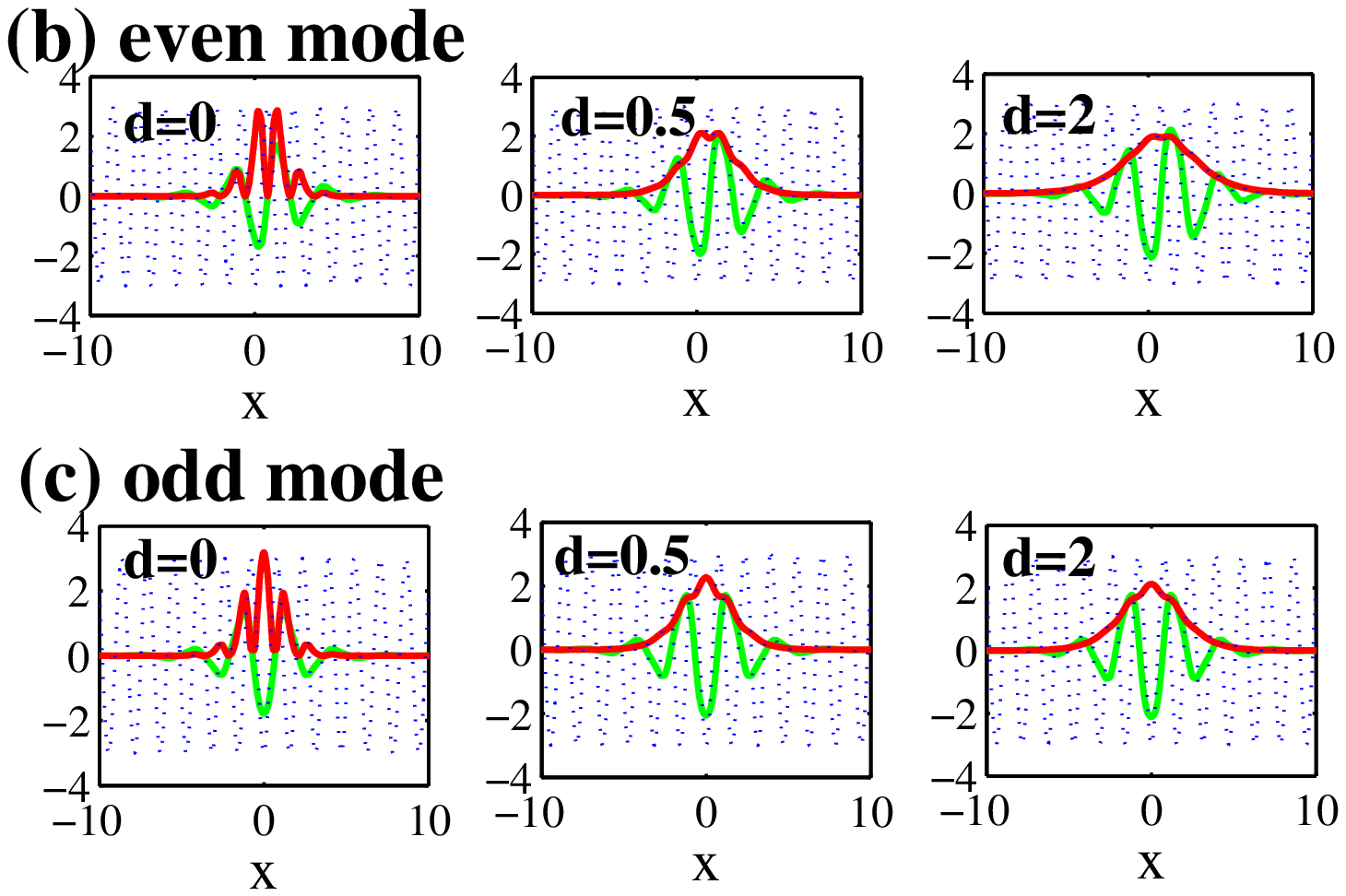}
\caption{(a)Families of even-modes (red) and odd-modes (blue) of gap solitons for different non-locality in the internal reflection and the first Bragg gap regions.
Field profiles of even-modes (b) and odd-modes (c) in the first Bragg gap region for different strength of non-localities, $d = 0, 0.5$ and $2$ are shown with a fixed wave number vector $b = -2$.
The green, red, and blue lines are the field $u(x)$, nonlinear index modulation $n(x)$, and periodical potential $V(x)$, respectively.}
\label{Fig:F3}
\end{figure}
\subsection{Bright gap soliton in nonlocal medium}
By applying the Bloch solution near the band edge of linear eigenvalue problem as an initial trial solution, we find different families of bright gap solitons $\Psi(x,z)= u(x)e^{ibz} $ numerically with a standard relaxation technique and the boundary conditions $u(\pm \infty)\approx 0 $.
Families of nonlinear gap soliton solutions can be found, as shown in Fig. \ref{Fig:F3}(a).
Here the bifurcation curves for gap solitons of even-mode and odd-mode in the internal reflection and the first Bragg gap regions through the relations of $b$ and soliton power $U \equiv \int |\Psi(x)|^2 d\,x$ are shown in red and blue lines, respectively. 
The two distinct types of solitons, on-site (even-mode) and off-site (odd-mode), are defined by their relative position of the center of wave functions with respect to external periodic potential \cite{Dabrowska}.
In other words, those centered on the minimum and the maximum potential of the lattice are classified as on-site and off-site.
Solutions of these gap solitons for on-site and off-site modes in the first Bragg gap region for different strength of non-localities, $d = 0, 0.5$, and $2$, are plotted in Fig. \ref{Fig:F3}(b, c) corresponding to the red and blue curves in Fig. \ref{Fig:F3}(a).

In comparison to the local nonlinear medium, $d = 0$ in Fig. \ref{Fig:F3}(a), as expected non-locality effect increases the formation power for gap solitons.
As a result, the gap soliton solutions, even- and odd-modes both, have a broader width of profile and a smaller level of amplitude.
In addition to the localized wave packet, gap solitons also have similar oscillatory tails as linear Bloch modes in the bands.
The oscillatory tails are significant in particular near the edge of linear band.
Compared to the solitons in the first Bragg gap region, gap soliton solutions in the internal reflection region have smooth envelope function and no oscillation tails are observed \cite{Torner-gap}. 
Additionally, the nonlinear refractive index distribution $n(x)$ in the internal reflection region is a smooth symmetric bell-like shape without pronounced local maximums on top of it.
In the following, we would show that the oscillation nature of Bloch wave makes solitons in the Bragg gap regions exhibit better stability and higher mobility. 

\subsection{Stability of gap solitons}
\begin{figure}
\includegraphics[width=8.0cm]{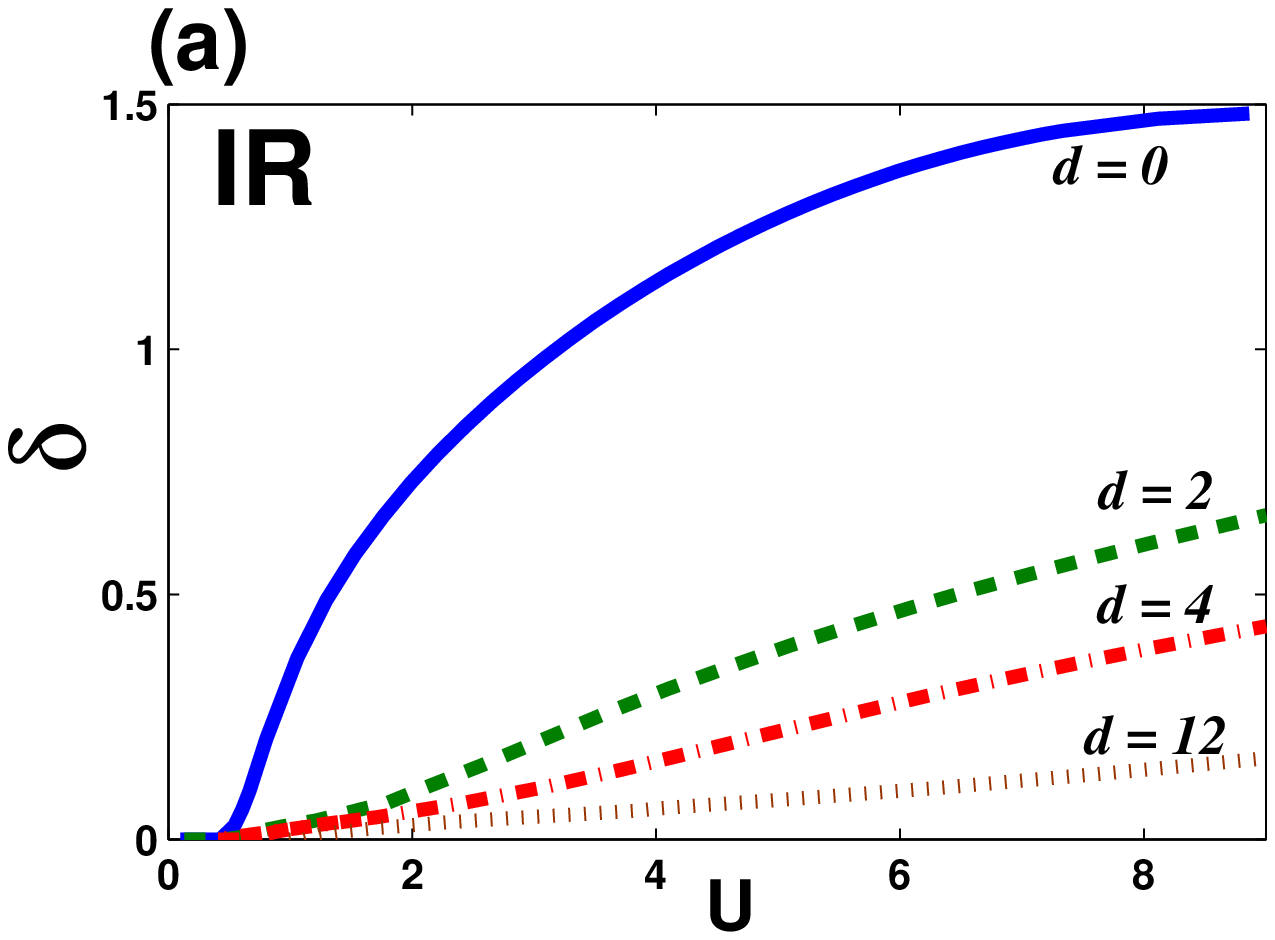}
\includegraphics[width=8.0cm]{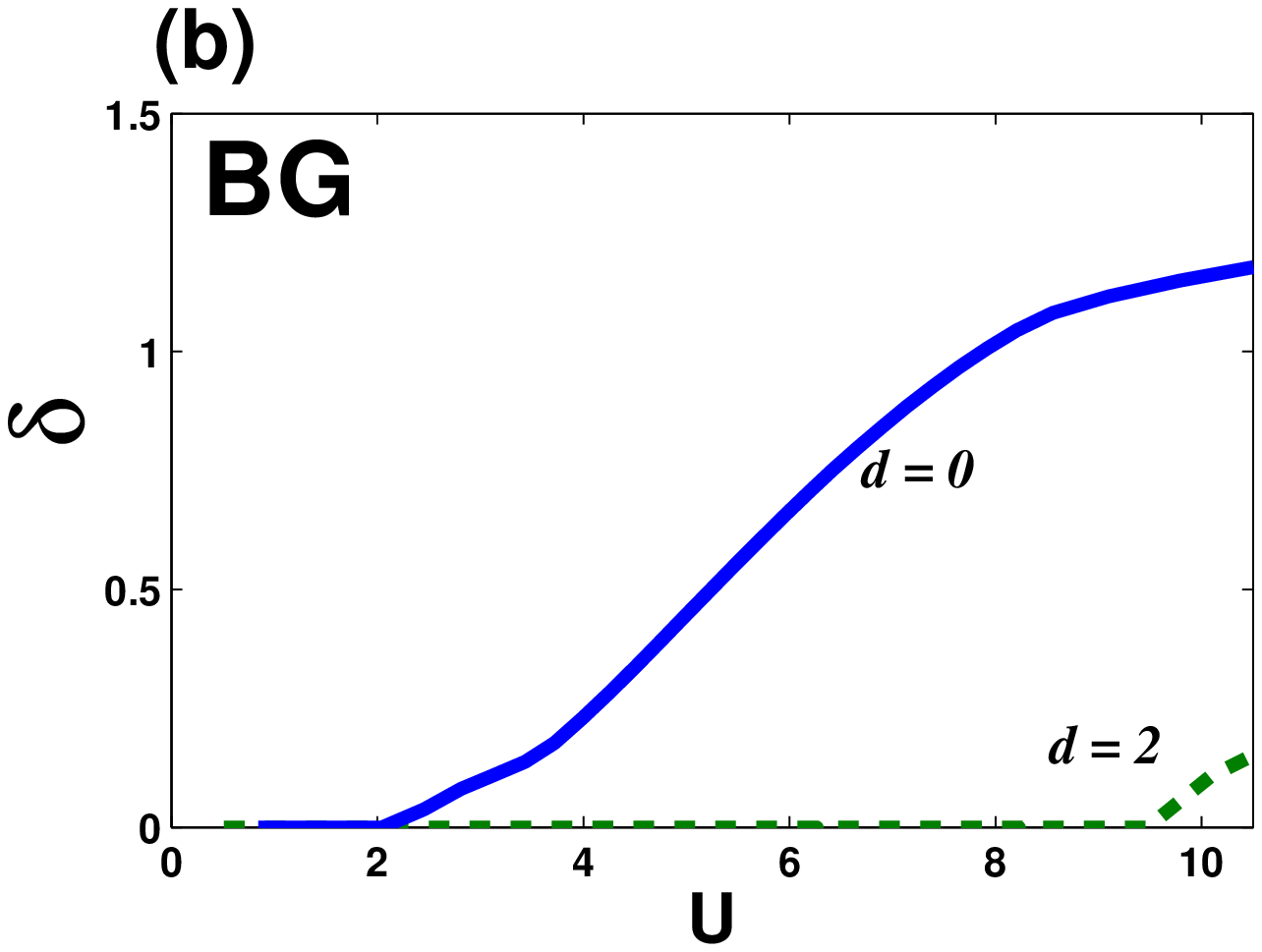}
\caption{Modulation instability of gap soliton with different non-localities, $d$. The growth rates of the small perturbation $Im(\delta)$ in the internal reflection (IR) and the first Bragg gap (BG) regions are shown in (a) and (b), respectively.}
\label{Fig:F4}
\end{figure}

The stability of gap soliton solutions is calculated through linear stability analysis, with the perturbed gap soliton solutions,
\begin{eqnarray}
&&u = u_0(x)e^{ibz} + \epsilon[p(x)e^{i\delta z} + q(x)e^{-i\delta^{*}z}]e^{ibz},\label{EqSLu}\\
&&n  = n_0 + \Delta n \label{EqSLn}
\end{eqnarray}
where $\epsilon\ll 1$, $u_0(x)$ is the unperturbed solution, and $Im\{\delta \}$ indicates the growth rate of the perturbations.
We linearize Eq.(\ref{EqSLu},\ref{EqSLn}) around the stationary solution and obtain, to the first order in $\epsilon$, the linear eigenvalue problem for the perturbation modes:
\begin{equation}
	\left(
	\begin{array}{ccc}
		\bf \hat{L}_0 + \hat{N}_0	&
		\bf \hat{N}_0\\
		\bf -\hat{N}_0	&
		\bf -\hat{L}_0 - \hat{N}_0
	\end{array}
	\right)
	\left(
	\begin{array}{clr}
		p\\
		q
	\end{array}
	\right)
	= \delta
	\left(
	\begin{array}{clr}
		p\\
		q
	\end{array}
	\right)
\label{EqSM}
\end{equation}
where
\[
\bf{ \hat{L}_0}\equiv n_0 - V(x) - b + \frac{1}{2} \frac{\partial^2}{\partial x^2},
\]
and 
\[
\bf{\hat{N}_0}\equiv |u_0|^2 (\bf{1}-d\frac{\partial^2}{\partial x^2})^{-1}.
\]
We solve the eigenvalue problem in Eq.(\ref{EqSM}) for unstable eigen functions with imaginary or complex propagation constant $\delta$ by standard matrix eigenvalue packages.
As the case of local nonlinear photonic crystals, it turns out that the on-site (even-mode) soliton family is modulationally stable while the off-site (odd-mode) soliton family is not \cite{Pelinovsky2004}.
Figure \ref{Fig:F4}(a,b) illustrate that nonlocal effect significantly reduces the growth rate of unstable spectral mode for off-site solitons in the internal reflection \cite{Torner-gap} as well as the first Bragg gap regions.
Furthermore, off-site gap solitons in the first Bragg gap region suffer stronger suppression of instability due to the diffusion of refractive index than the case in the internal reflection region.

As seen in Fig. \ref{Fig:F4} clearly, with a small strength of non-locality the modulation instability of gap solitons in Bragg gap region is impressively suppressed.
A simple reason is that nonlocal effect can smooth over the oscillation tails of refractive index $n$ due to the diffusion mechanism.
In other words, the smoothness of refractive index $n$ indicates that the effective potential $V_{eff}=V(x) - n(x)$  turns more broadened.
Therefore, as the strength of long range interaction increases, soliton solutions become more stable due to the broadening effect of the effective potential, especially in the Bragg gap region. 
Because gap soliton solutions in the Bragg gap regions access oscillation tails and their profiles extend more apparently in space, in such a way the nonlocal effect become more apparently to smooth out the effective potential and to stabilize solitons.
\begin{figure}
\includegraphics[width=7.0cm]{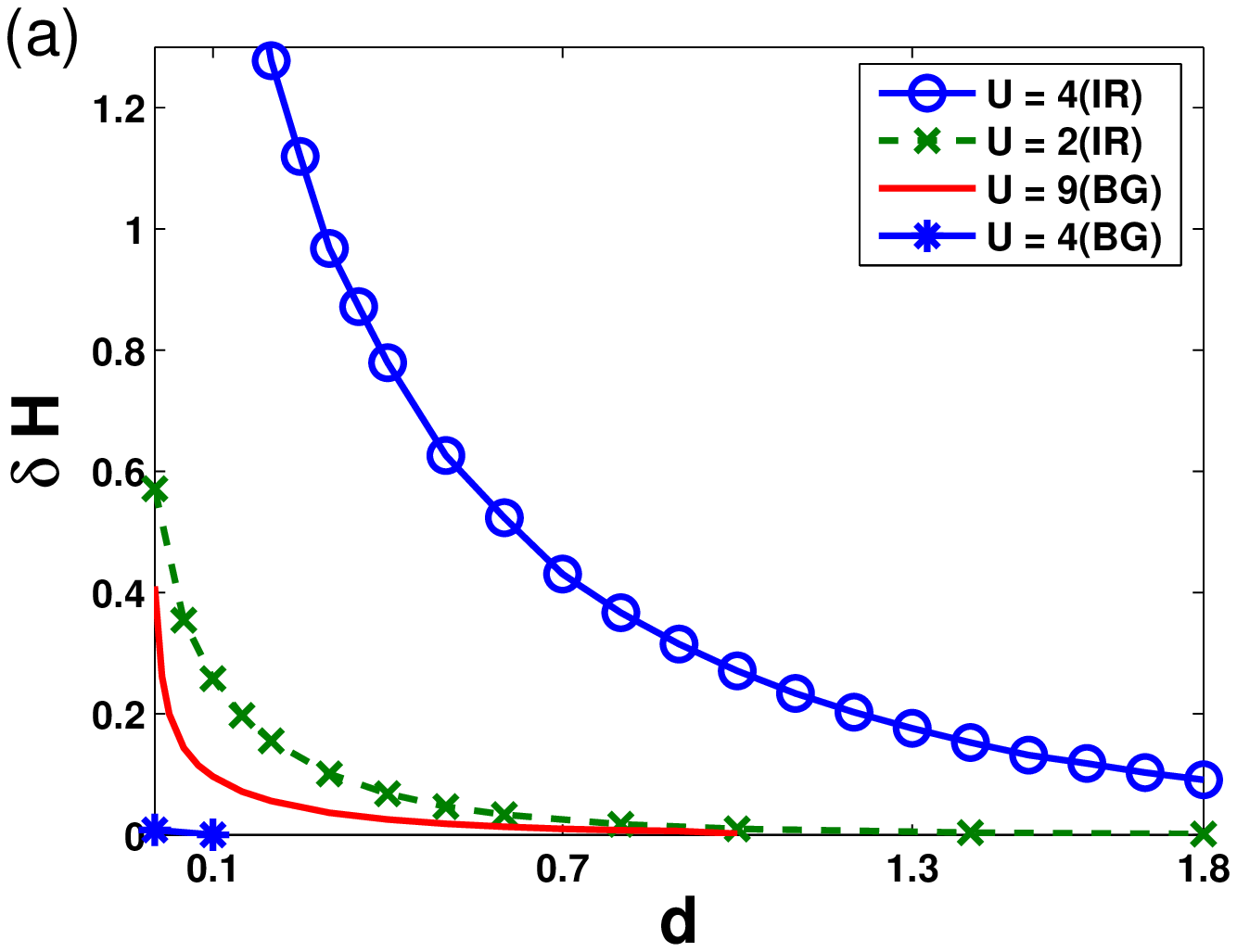}
\includegraphics[width=7.0cm]{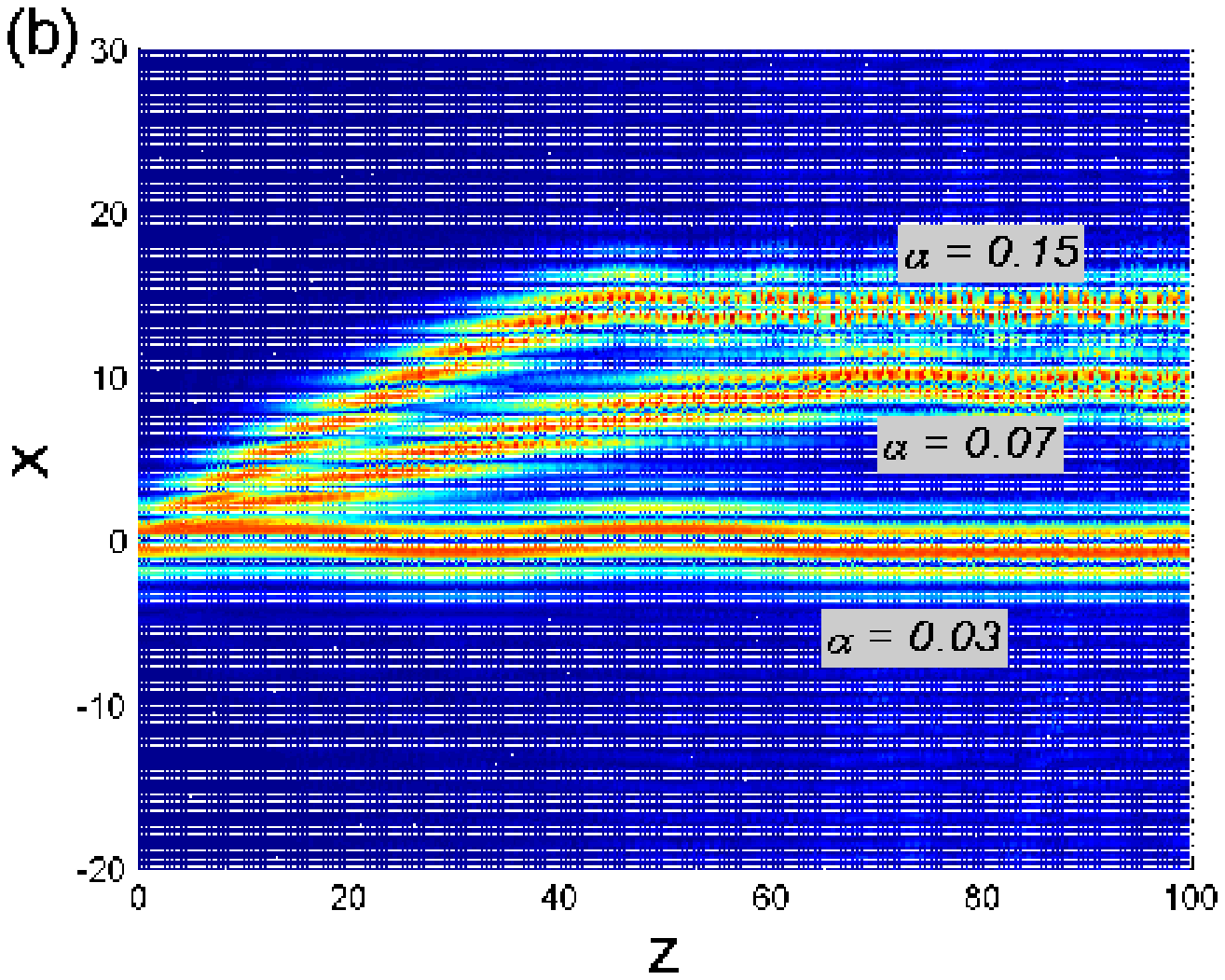}
\caption{(a) PN potential height ($\delta$ H) v.s. different non-localities ($d$) with a comparison of gap solitons in the internal reflection (IR) and the first Bragg gap (BG) regions. (b) Gap solitons propagation trajectories in Bragg gap (BG) region with fixed $d=0.5$ and $U=9$ but different initial kinetic energies ($\alpha = 0.03, 0.07$, and $0.15$).}
\label{Fig:F5}
\end{figure}
\subsection{Mobility and collision of gap solitons}
In this section we study the mobility of these gap soliton solutions by calculating their Peierls-Nabarro (PN) potential barrier which is introduced as the height of an effective periodic potential generated by the lattice discreteness. 
PN potential barrier defines the minimum energy required to move the center of mass of a localized wave packet by one lattice site \cite{Kivshar93}. 
Extending this definition to the continuous system, we can define PN potential barrier as the difference of system Hamiltonian between even-modes (on-site) and odd-modes (off-site), i.e.
\begin{eqnarray}
&&\delta H = H_{even} - H_{odd} \label{EqdH}\\
&&H = \int^{\infty}_{-\infty} \left[\left|{\frac{\partial{u}}{\partial x}}\right|^2 - \frac{1}{2}\left|u\right|^2n\right]dx. \label{EqH}
\end{eqnarray}

\begin{figure}[t]
\includegraphics[width=8.0cm]{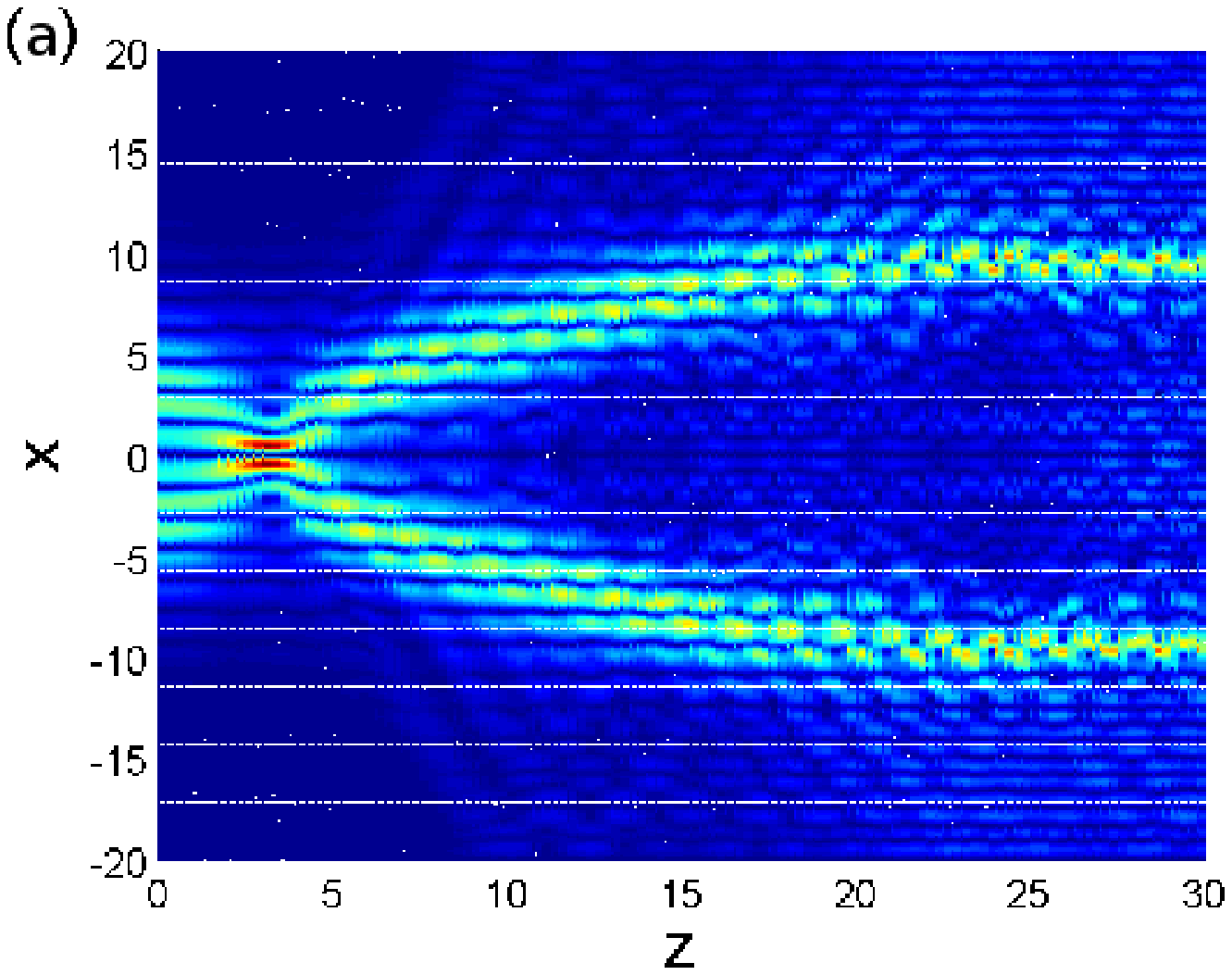}
\includegraphics[width=8.0cm]{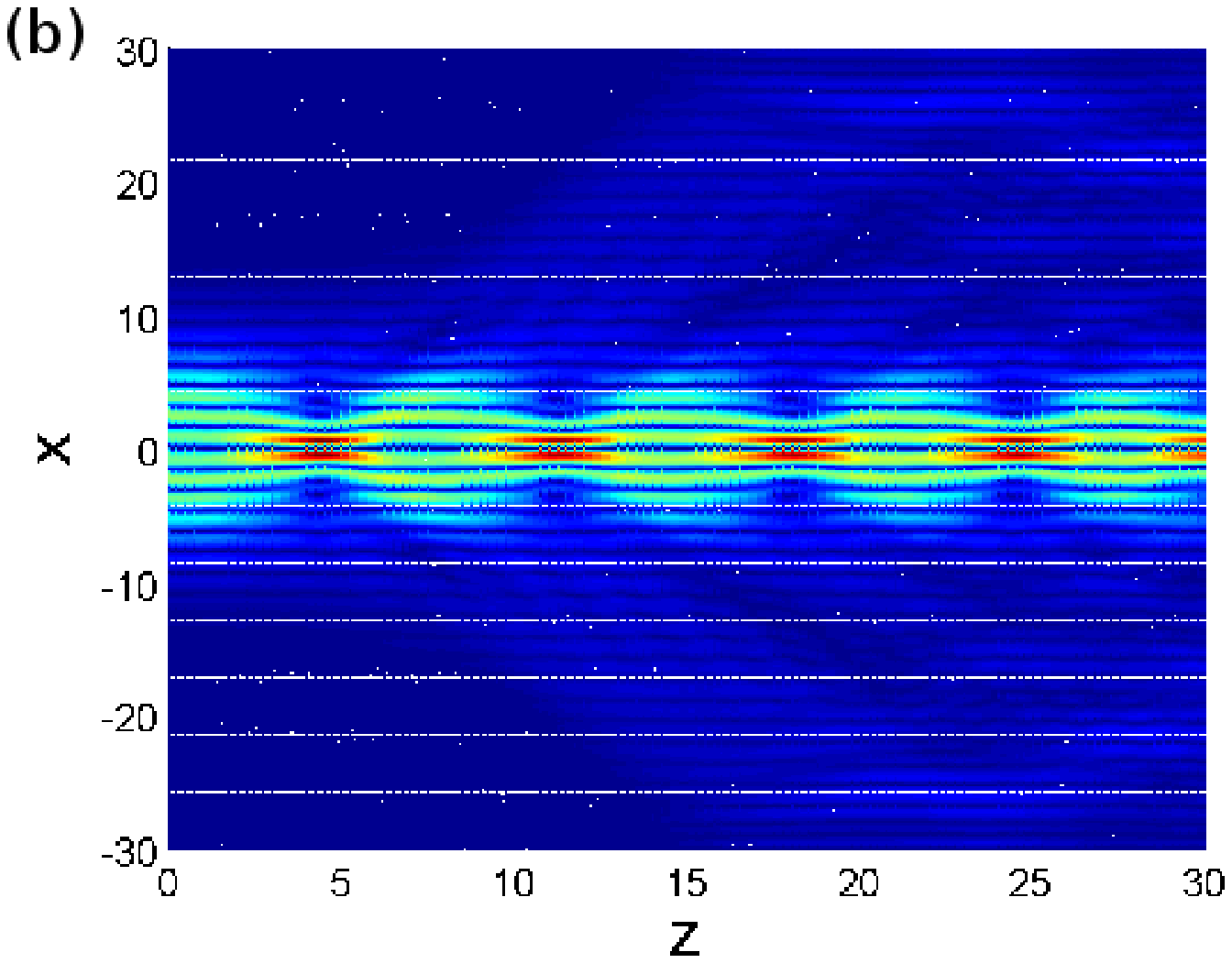}
\caption{Collision of two solitons in photonic crystals with local (a) and nonlocal (b) non-linearities. The degree of non-locality is set as $d=0.5$.}
\label{Fig:F6}
\end{figure}
Consequently, PN potential states the smallest amount of energy that a gap solitons needed to gain in order to start moving along the lattice.
It is clearly seen that in the first Bragg gap region the PN potential barrier is drastically reduced in comparison to the local nonlinearity, $d = 0$, as well as the internal reflection region, as shown in  Fig. \ref{Fig:F5}(a).
With the oscillation tails similar to the linear Bloch waves, nonlocal solitons in the Bragg gap regions are more stable and more movable than those in the internal reflection band \cite{Torner-gap}.
The higher the barrier $\delta H$, the larger the incident kinetic energy required to overcome the barrier. 
The reduction of the PN barrier is confirmed by numerical simulations of Eq.(\ref{eqGP1}) for gap solitons with fixed power but different initial kinetic energies.
The initial conditions for Fig. \ref{Fig:F5}(b) are set with $ u(x,z=0) = u_{0} e^{i\alpha x}$, where $u_0$ is the stationary solution in Bragg gap region and $\alpha$ stands for transverse momentum at incidence.
When soliton crosses the lattice it radiates and loses energy.  
Eventually these gap solitons are captured by one of the lattice channel.
However, while the nonlocality effect comes to play, the total effective potential is shallowed by long range interaction. 
These nonlocal gap solitons are more free to move in the transverse direction due to a lower PN potential height.

Based on the results of the PN potential barrier reduction for solitons in the Bragg gap regions with non-locality, we demonstrate a potential-free collisions between two gap solitons within the photonic crystals.
It is well known that without periodic potentials, solitons experience elastic collisions when the relative phase between them is in-phase.
The elastic collision between two solitons is destructed in the photonic systems due to the confinement of the PN potential barrier, as shown in Fig.\ref{Fig:F6}(a) for the local case with $d = 0$.
But with nonlocal nonlinear response, in Fig.\ref{Fig:F6}(b) we show that gap solitons can revive an elastic-like collision even in the photonic systems.
In principle, the collisions between gap solitons are very complicated and unpredicted due to the interplays among nonlinearity, periodic potential height, and dispersion/diffraction effects.
Nonetheless, in this simulation, only with a small strength of non-locality, $d = 0.5$ one can have a clear picture of gap soliton interactions.

\section{Conclusion}
In conclusion, we demonstrate the existence of gap soliton solutions in nonlocal nonlinear photonic crystals in the internal reflection and Bragg gap regions.
The stability and mobility of such novel gap solitons are calculated through the linear stability method and the Peierls-Nabarro potential barrier.
Compared to the internal reﬂection region, nonlocal gap solitons in the Bragg gap regions become not only more stable but also more movable due to the oscillation tails of wave packets.
Moreover we reveal that it is possible to have elastic-like collisions between gap solitons with the help of a small strength of nonlocal effect.
Supported by the result of this study and current technology about controllable nonlocal nonlinear media, such as photorefractive crystals, nematic liquid crystals and thermo-optical materials, we believe the results in this work should provide a new way for soliton-based photonic devices.

Authors are indebted to Tsin-Fu Jiang and Yinchieh Lai for useful discussions.
This work is supported by the National Science Council of Taiwan with the contrast number 95-2112-M-007-058-MY3 and NSC-95-2120-M-001-006.

\end{document}